\documentclass[epj]{svjour}
\usepackage{graphics}
\usepackage{amsmath}
\usepackage{graphicx,bm,times,mathptmx}
\begin{document}
\title{A Check-up for the Statistical Parton Model}

\author{Franco Buccella\inst{1}\thanks{buccella@na.infn.it} \and Sozha Sohaily\inst{2}
}
%
%
\institute{INFN, Sezione di Napoli, Via Cintia, I-80126Napoli, Italy \and Faculty of Physics, Shahid Bahonar University of Kerman, Kerman, Iran}
\date{Received: date / Revised version: date}
%
\abstract{
We compare the parton distributions deduced in the framework
of a quantum statistical approach for both the longitudinal
and transverse degrees of freedom with the unpolarized distributions
measured at Hera  and with the polarized ones proposed in a
previous paper, which have been shown to be in very good agreement
also with the results  of experiments performed after that
proposal.
The agreement with Hera data in correspondence of very similar
values for the "temperature" and the "potentials" found in the
previous work gives a robust confirm of the statistical model.
The feature of describing both unpolarized and polarized parton
distributions in terms of few parameters fixed by data with large statistics and small systematic errors makes very attractive the parametrization proposed here.
} 

%
\maketitle
\section{Introduction}
About twenty years ago [1] the similar shapes of the
polarized structure function $x g^p_1(x)$ and of the difference $F^p_2(x)-F^n_2(x)$ suggested
that for the shapes and the first moments of the valence quark partron
distributions there is the correlation dictated by Pauli principle, which
also accounts for the defect in the Gottfried sum rule [2] found
experimentally [3], related to the isospin asymmetry
$\bar{d}-\bar{u}$ advocated since a long time [4].
The role of Pauli principle is a robust motivation to write for the distributions of the valence partons Fermi-Dirac functions in the
variable $x$, which appears in the parton model sum rules.

\let\thefootnote*\footnotetext{We thank Prof. Sivers for the remark that the role
of Pauli principle is a consequence of confinement.}

The shape-first moment correlation implied by Pauli principle accounts for
the dramatic high $x$ decrease of the ratio $F^n_2(x)/F^p_2(x)$ and
for the increasing (decreasing) $x$-dependance of the positive (negative)
ratio $\Delta{u(x)}/u(x)$ ($\Delta{d(x)}/d(x)$) [5].
After many attempts a satisfactory description of a selected choice of precise
unpolarized and polarized structure functions has been obtained [6] by
adding to a Fermi-Dirac expressions for the quark partons an unpolarized
and iso-scalar term, which may be interpreted as the gluon contribution at
order $\alpha_s$, with a power in front of the FD functions multiplied by a constant proportional to
the "potential" associated to each parton for the valence quarks.
A crucial role to fix the shape of the gluon distribution (evidently a
Bose-Einstein function in the $x$ variable) is plaid by the equilibrium
conditions for the elementary QCD processes (emission of a gluon by a quark
and conversion of a gluon in a $q \bar{q}$ pair), which imply a zero value
for the potentials of the gluons with both helicities ( Bose-Einstein
turns into Planck and $\Delta{G(x)}=0 $) and opposite values for the valence
quarks and their antiparticles with opposite helicity.
In [6], we have been able to describe both unpolarized and
polarized distributions in terms of few parameters and in subsequent works,
we compared our predictions with experimental results obtained in the
following years [7, 8] in agreement both for the
polarized structure functions $g^{p,n,d}_1$ and for the unpolarized
structure functions measured in the electromagnetic and weak DIS at Hera.
Also we tried to explain the "ad hoc" factors $X^h_q$, we had to
introduce for the non-diffractive part of the fermion parton distributions.
We realized that with the extension of the statistical approach to the
transverse degrees of freedom from a sum rule for the transverse energy
one is able to predict a gaussian dependance on $p_T$ with a width
increasing with $x$ and fix the "transverse potentials" to reproduce the
"ad hoc" factors introduced for the valence partons in [6].
The purpose of this paper is to provide a general check of the approach
proposed in [6].
In the following section, we recall the expressions for the partons introduced
in [6]. In section 3, the consequences of the extension to the transverse
momenta are described, by keeping also into account of Melosh rotation [9].
In Section 4, our expressions are compared with the light parton distributions
obtained in the combined fit to $H_1-Zeus$ data performed at Hera [10]
for the unpolarized distributions and with the expressions for the polarized
distributions given in [6] and very successful to describe the
measurements depending on them after 2002 [7, 8, 11].
In Section 5, we give our conclusions.
\label{intro}

\section{The Statistical Model Proposed in 2002}

Let us recall the expressions and the values of the parameters,
which allowed to get a fair description of both unpolarized and polarized structure functions [6].
For the light valence partons at $Q^2=4 (GeV^2)$, we assumed:

\begin{equation}
x u^{+} (x) =\frac{A x^bX_{u^{+}}}{ exp[{(x -
X_{u^{+}})/\bar{x}}] + 1} +
\frac{ \tilde{A} x^{\tilde{b}}}{ exp({x/\bar{x}})+ 1}
\label{eq.1}
\end{equation}

and their antiparticles:

\begin{equation}
x \bar{u}^{+} (x) =\frac{\bar{A} x^{2b}(X_{u^{-}})^{-1}}{ exp[{(x+
X_{u^{-}})/\bar{x}}] + 1} +
\frac{ \tilde{A} x^{\tilde{b}}}{ exp({x/\bar{x}})+ 1}
\label{eq.2}
\end{equation}


and corresponding expressions for $u^{-}$, $d^{+}$, $d^{-}$ and their antiparticles with $X_{u^{+}}$ $\rightarrow$ $X_{u^-}$, $X_{d^+}$ and $X_{d^-}$, respectively for eq.(1), and
$X_{u^{-}} \rightarrow X_{u^{+}}$, $X_{d^{-}}$ and $X_{d^{+}}$
for eq.(2)

For gluons one assumed the Bose-Einstein form:

\begin{equation}
x G^{\pm}(x)=
\frac{A_g x^{b_g}}{exp{(x/\bar{x})}- 1}
\label{eq.3}
\end{equation}

The opposite values for the potentials $X_{u^{+}}$ and $X_{\bar{u}^{-}}$
and the vanishing  value  for $X_{G^{+}}$ and $X_{G^{-}}$,
which turns the Bose-Einstein into a Planck expression, follow from the
equilibrium condition with respect to the elementary QCD processes [12]:

\begin{equation}
q^{\pm}\rightarrow q^{\pm}+G
\label{eq.4}
\end{equation}

\begin{equation}
G \rightarrow q^{(\pm)}+\bar{q}^{(\mp)}
\label{eq.5}
\end{equation}



The coefficients in the first term of eq.1 were introduced to agree with
data and, just by guess, we assumed the same coefficient in the
denominator in eq.2.
The interpretation of the second term in eqs.(1) and (2) as the diffractive
contribution coming from the gluons lead to the assumption:

\begin{equation}
b_g =\tilde{b} + 1
\label{eq.6}
\end{equation}








The constraints:

\begin{subequations}
\begin{equation}
u - \bar{u} = 2
\end{equation}
\begin{equation}
d - \bar{d} = 1
\end{equation}
\label{eq.7}
\end{subequations}



and the requirement that the partons carry the longitudinal momentum of the proton fix $A$, $\bar{A}$ and $A_g$.

The 2002 fit was obtained with the following values of the parameters:

\begin{align}
&\bar{x} = 0.09907, X_{u^{+}} = 0.46128, X_{d^{-}} = 0.30174, X_{u^{-}} = 0.29766\nonumber\\&
X_{d^{+}} = 0.22775, A = 1.74938, \bar{A} = 1.90801, A_g =20.53\nonumber\\&
\tilde{A} = 0.08318, b = 0.40962, \tilde{b} = -0.25347
\label{eq.8}
\end{align}

\label{sec:1}

\section{The Extension of the Statistical Approach to the Transverse Degree of Freedom}

In [7], the predictions of the statistical model introduced
in [6] have been compared with measurements performed after with a particular success for
the polarized structure function $x g^n_1(x)$, expected to be negative at
small $x$ and positive at large $x$ and with a good precision for the
value of $x$, where it vanishes. In [8] a successful
test of the predictions has been performed for the unpolarized structure functions for
electromagnetic and weak DIS scattering at Hera.
Then to account for the "ad hoc" factors introduced for
the non-diffractive part of the valence parton distributions and for their
antiparticles, a sum rule has been assumed for the transverse energy [11],
which fixes in terms of a dimensional Lagrange multiplayer $1/\mu^2$
and of the transverse potentials, $\tilde{Y}^h_q$, the $p^2_T$ dependance,
which tends for large $p^2_T$ to a gaussian form, by other authors considered
without a justification, with a width proportional to $\sqrt{x}$.

The equations for the parton distributions depending on
$x$ and $p^2_T$ for the non-diffractive part of the unpolarized distributions are:

\begin{footnotesize}
\begin{align}
&xu(x,p^2_T)=\frac{1}{x\mu^2}(\frac{A^{\prime}x^b}{e^{(x-\tilde{X}^{+}_u)/\bar{x}}+1}\frac{1}{e^{(2p^2_T/\mu^2(x+\sqrt{x^2+(p^2_T/P^2_z)}))-\tilde{Y}^{+}_u}+1}\nonumber\\&
+\frac{A^{\prime}x^b}{e^{(x-\tilde{X}^{-}_u)/\bar{x}}+1}\frac{1}{e^{(2p^2_T/\mu^2(x+\sqrt{x^2+(p^2_T/P^2_z)}))-\tilde{Y}^{-}_u}+1})
\label{eq.9}
\end{align}
\end{footnotesize}

\begin{footnotesize}
\begin{align}
&xd(x,p^2_T)=\frac{1}{x\mu^2}(\frac{A^{\prime}x^b}{e^{(x-\tilde{X}^{+}_d)/\bar{x}}+1}\frac{1}{e^{(2p^2_T/\mu^2(x+\sqrt{x^2+(p^2_T/P^2_z)}))-\tilde{Y}^{+}_d}+1}\nonumber\\&
+\frac{A^{\prime}x^b}{e^{(x-\tilde{X}^{-}_d)/\bar{x}}+1}\frac{1}{e^{(2p^2_T/\mu^2(x+\sqrt{x^2+(p^2_T/P^2_z)}))-\tilde{Y}^{-}_d}+1})
\label{eq.10}
\end{align}
\end{footnotesize}

and similar expressions for $x\Delta{u}$ and $x\Delta{d}$ except the minus sign  instead of the positive sign between the two terms in the right hand side of eqs.(9) and (10) and also, the factor coming from Melosh transformation.

For the antiquarks, one has the expressions in which the $\tilde{X}$'s and the $\tilde{Y}$'s have opposite sign than their antiparticles with opposite and a different normalization constant, $\bar{A}$ helicity.
For the gluons, we write the expression:

\begin{equation}
x G(x)=\frac{A_g x^{b_g}e^{-C_g(x+c_g)^2}}{e^{(x/\bar{x})}-1}\left(1+ln(\frac{x+r}{x+u})\right)
\label{eq.11}
\end{equation}

where the two slowly varying factors, which slightly modify the behavior
dictated by the Planck expression for gluons, are needed to agree almost
perfectly with [10].

By integrating in $d p^2_T$ eqs.(9) and (10) one gets:

\begin{align}
&xq(x)=\frac{A^{\prime}x^b}{e^{(x-\tilde{X}_q^+)/\bar{x}}}\left[ln(1+e^{\tilde{Y}_q^+})+\frac{2\mu^2(1-x)}{Q^2}(-Li_2(-e^{\tilde{Y}_q^+}))\right]\nonumber\\&
+\frac{A^{\prime}x^b}{e^{(x-\tilde{X}_q^-)/\bar{x}}}\left[ln(1+e^{\tilde{Y}_q^-})+\frac{2\mu^2(1-x)}{Q^2}(-Li_2(-e^{\tilde{Y}_q^-}))\right]
\label{eq.12}
\end{align}

\begin{align}
x\Delta{q(x)}=\frac{A^{\prime}x^b}{e^{(x-\tilde{X}_q^+)/\bar{x}}}\left[ln(1+e^{\tilde{Y}_q^+})\right]
-\frac{A^{\prime}x^b}{e^{(x-\tilde{X}_q^-)/\bar{x}}}\left[ln(1+e^{\tilde{Y}_q^-})\right]
\label{eq.13}
\end{align}

Where $q = u, d$ and the small correction proportional to the ratio $2\mu^2/Q^2$ is absent for the polarized distributions since it is exactly compensated by the consequence of the Melosh-Wigner rotation [9].

In the limit of large $Q^2$ we get the factor
$\emph{ln}{(1 + exp[\tilde{Y}^h_q])}$, which we can reasonably assume
to be proportional to $\tilde{X}^h_q$ (valence quark with larger first moments are
expected to have broader shapes both in the longitudinal and transverse
degrees of freedom). While with the above assumption we can get the "ad
hoc" factors for valence partons, for their antiparticles a similar
property can be obtained only approximately.
In fact, while we guessed in [6] the product of the factors for the
valence quark of given helicity and of their antiparticle with opposite
helicity to be constant, the function:

\begin{equation}
\emph{ln}{(1 + exp{[\tilde{Y}]})}\emph{ln}{(1 + exp[-\tilde{Y}])}\nonumber\\
\end{equation}

has its maximum $(\ln{2})^2$ at $\tilde{Y}=0$ , and therefore the assumption of the
proportionality between $\emph{ln}{(1 + exp[\tilde{Y}^h_q])}$ and $\tilde{X}^h_q$
implies a smaller coefficient for the more rare light antipartons with the consequence
that with the same $\tilde{X}^h_q$ for the non-diffractive part the ratio
$\Delta \bar{u}/\bar{u}$  becomes more positive
and $\Delta \bar{d}/\bar{d}$ more negative.

If we apply to the gluons the extension to the transverse degrees of freedom,
with a vanishing transverse potential for both helicities, we find a divergent
expression, but we can avoid the inconvenient by observing that with $b_g = 1$,
as we impose according to the radiation interpretation of the gluon component,
its contribution to the longitudinal and transverse energy sum rules, albeit
divergent, are in the ratio $\mu^2/\bar{x}$. Since for the gluon we did not
introduce, as for the fermionic partons, arbitrary factors, we can take for the gluon
distribution the form assumed in eq.(11) with $b_g = 1$ and assume for the
contribution to the transverse energy [11] of the gluon the value given by the product
of its contribution to the longitudinal sum rule times $\mu^2/\bar{x}$.

Another difference with respect to [6] is that the Melosh rotation
[9] for the polarized distributions implies a very small difference
for the normalization of the polarized and unpolarized distributions.

\label{sec:2}

\section{The Comparison with Hera Data for the Statistical Parton Model}

We want to test the statistical parton model by describing with the form
given in the previous section for the unpolarized light fermion partons with the
result of the combined $H_1-ZEUS$ fit [10]. As long as the polarized
distributions of the light partons, the success in describing the
$x g^{p,n,d}_1$ structure functions and the confirm
[11] of our prediction of a positive value for $\Delta{\bar{u}}$ and a
negative for $\Delta{\bar{d}}$ gives us the motivation to require that the
polarized distributions are just the ones found in [6].
With respect to [6] article we assume the same exponent, $b$, also for
the non-diffractive part of the light anti-quarks and we fix $b_g$ to
be $1$, which is the value expected for a radiation term.
The extension to the transverse degrees of freedom implies  a small
difference for the unpolarized and polarized distributions as a
consequence of the Melosh [9] rotation.
Also, the proportionality between:

$\emph{ln}{(1 + \exp{[\tilde{Y}_q}])}$ and $\tilde{X}_q$ does not imply the
proportionality between $\emph{ln}{(1 + \exp{[\tilde{-Y}_q}])}$
and $1/\tilde{X}_q$ with the consequence that for the non-diffractive part of the antiquarks
with $\tilde{Y}^h_q$ given by eq.11 one has a more negative value for
$\Delta{\bar{d}}(x)/\bar{d}(x)$ and a more positive value for
$\Delta{\bar{u}}(x)/\bar{u}(x)$.
In conclusion the formulas have the following differences from [6]:

i) Instead of the "ad hoc" factors $\tilde{X}^h_q$ for valence quarks we have
the factors $\emph{ln}{(1 + exp[\tilde{Y}^h_q])}$ depending on the transverse potentials,
while for their antiparticles with opposite helicity we have the factors
$\emph{ln}{(1 + exp[-\tilde{Y}^h_q])}$.

ii) By keeping into account of the Melosh rotation [9], we have
the difference between the polarized and the unpolarized distribution
with the additional factor for these proportional to $\frac{2\mu^2(1-x)}{Q^2}$,
which anyway gives a small correction, since $\mu^2$ is about $0.1
(GeV)^2 $ and $Q^2=4(GeV)^2$.

iii) We take the same exponent, $b$, for the power for the light
anti-quarks, and fix to $b_g=1$, the exponent for the gluon, as it is
appropriate for a radiation term.

\begin{figure}[h]
\centering
\resizebox{0.48\hsize}{!}
{\includegraphics*{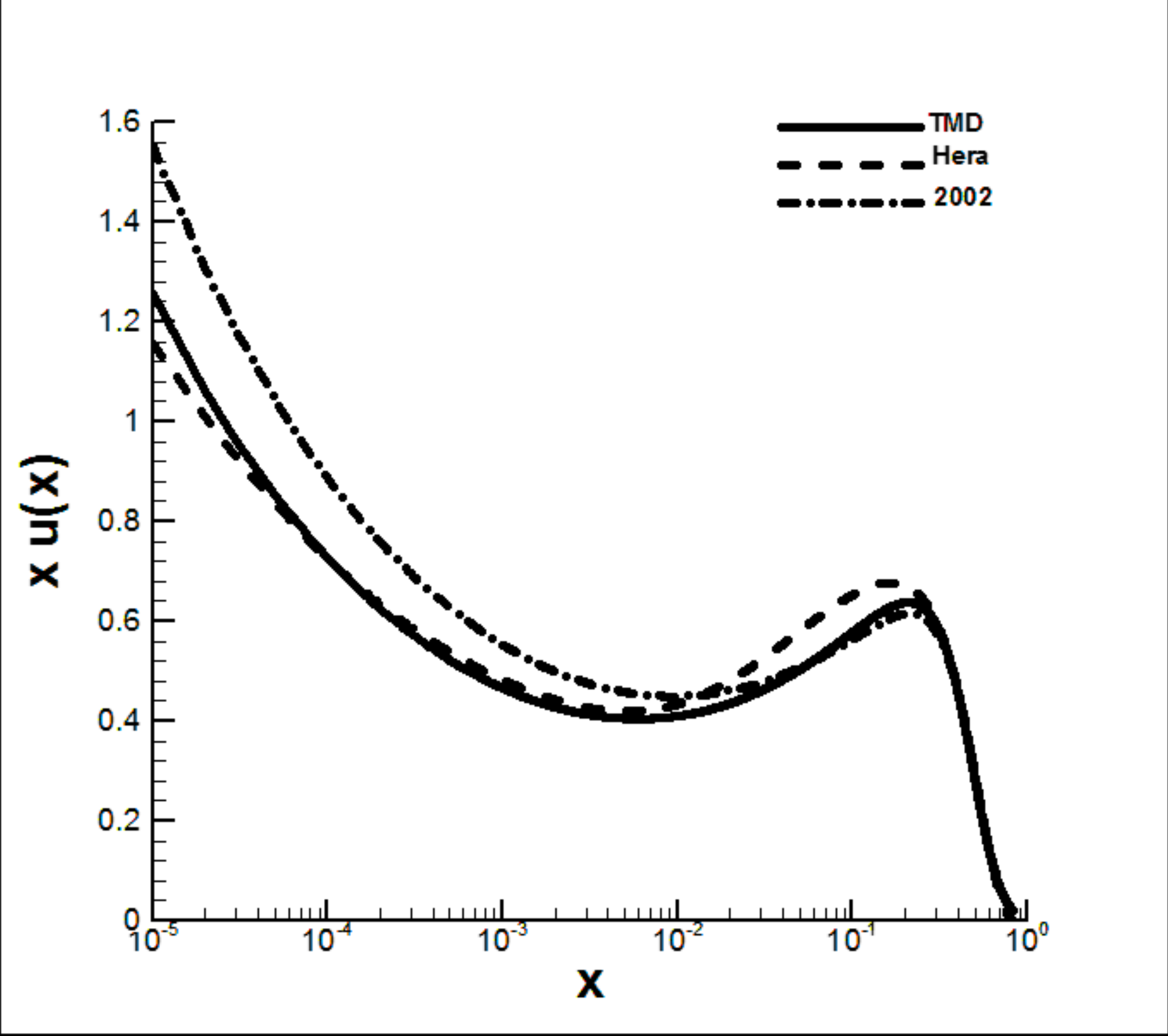}}
\resizebox{0.48\hsize}{!}
{\includegraphics*{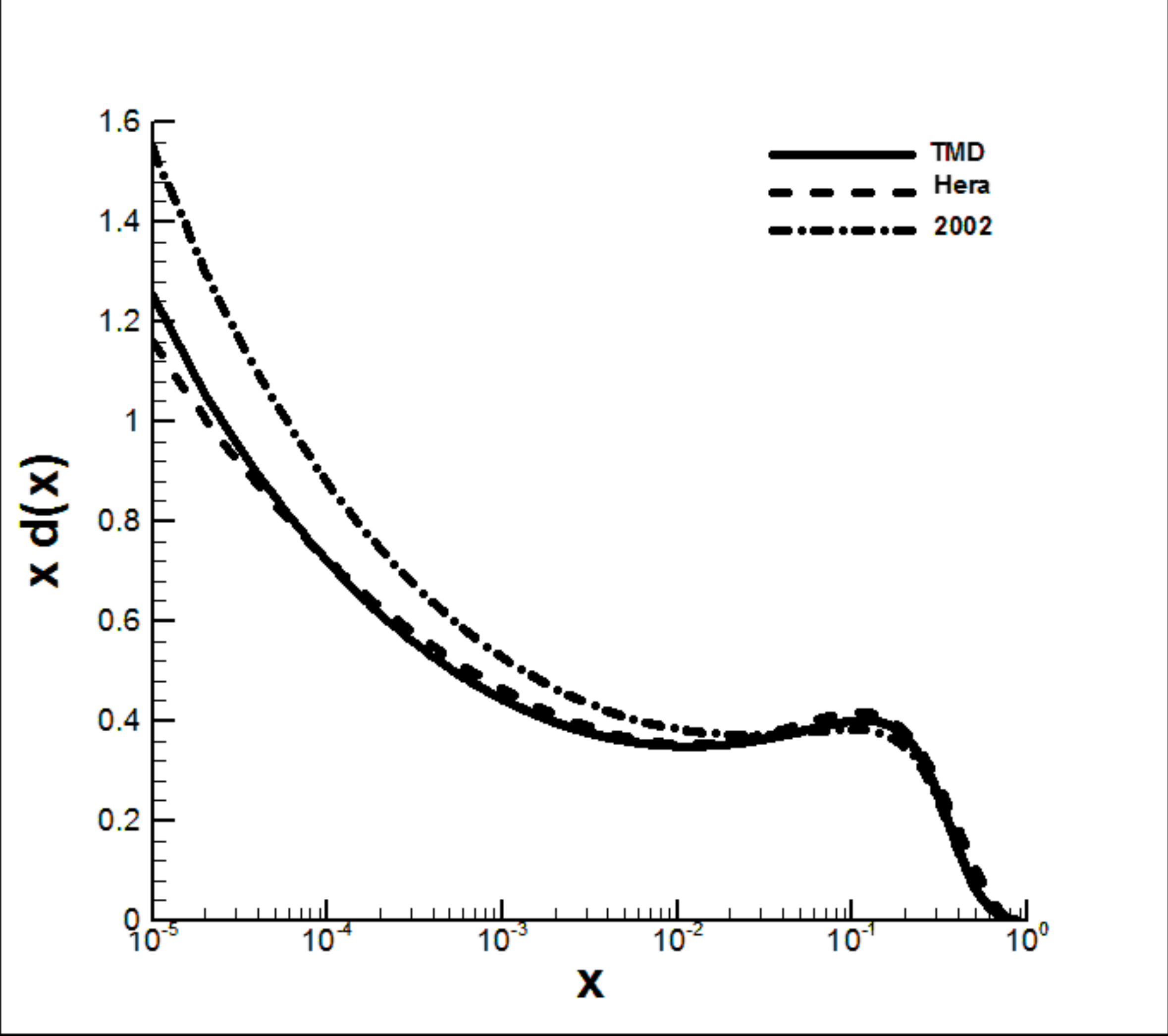}}
\caption{The distribution of quarks up and down in comparison to the 2002 results and Hera data.}
\label{fig1}
\end{figure}

\begin{figure}[h]
\centering
\resizebox{0.48\hsize}{!}
{\includegraphics*{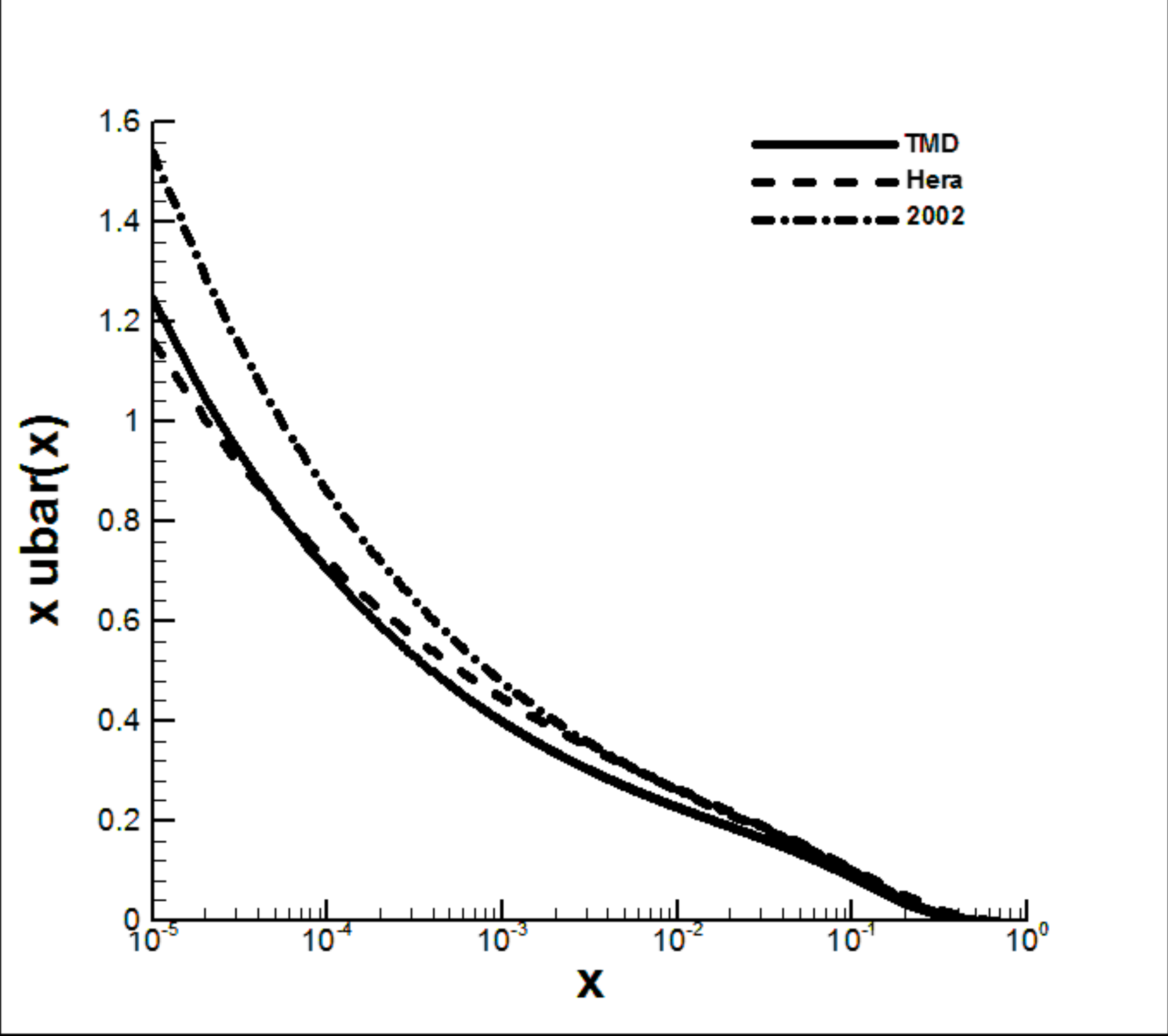}}
\resizebox{0.48\hsize}{!}
{\includegraphics*{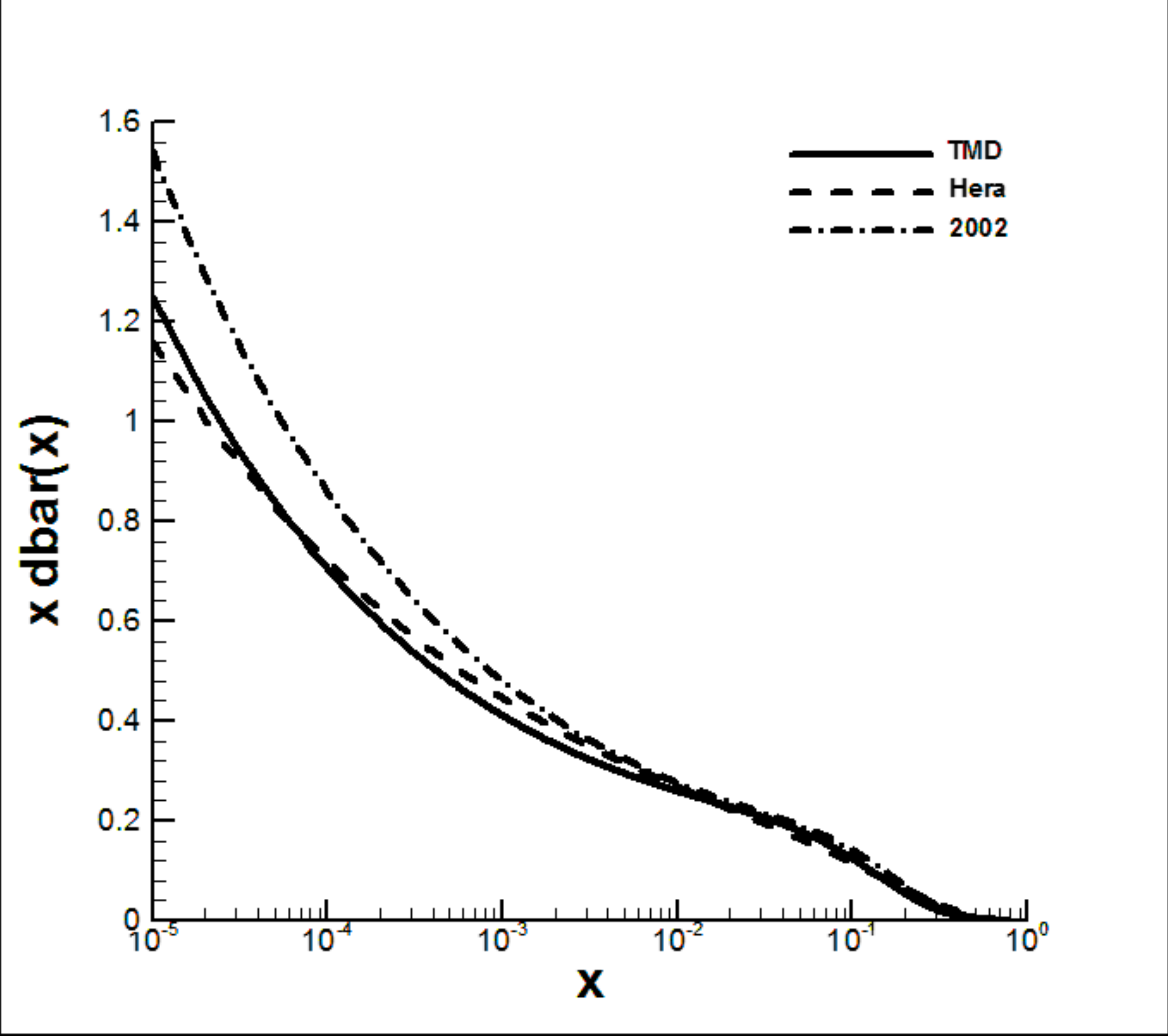}}
\caption{The distribution of anti- quarks upbar and downbar in comparison to the 2002 results and Hera data.}
\label{fig2}
\end{figure}

In figures (1) and (2), we compare the expressions given in the previous section with the determination of HERA through
the combined fit $H_1-ZEUS$ of the light valence partons and their
antiparticles. The success in the comparison with experiment of the polarized structure
functions [7, 8] and the evidence for a positive
$\Delta{\bar{u}}(x)$ and for a negative $\Delta{\bar{d}}(x)$ [11] also
quantitatively as predicted in [6] motivates our demand that the
polarized distributions of the light partons be equal to the ones proposed
in that paper.
We compare in figures 3-4 the polarized distributions with the expressions found in [6]. Finally we fix the parameters in eq (11) to agree with [10],
as shown in figure 5.
The values for the parameters are the following:

\begin{align}
& \tilde{X}_{u^{+}} = 0.446, \tilde{X}_{d^{+}} = 0.222, \tilde{X}_{d^{-}} = 0.320, \tilde{X}_{u^{-}} = 0.297\nonumber\\&
\tilde{Y}_{u^{+}} = 1.050, \tilde{Y}_{d^{+}} = 0.01, \tilde{Y}_{d^{-}} = 0.360, \tilde{Y}_{u^{-}} = 0.293\nonumber\\&
A^{\prime} = 0.615, \bar{A}^{\prime} = 3.50, \mu^2= 0.0938, b =\bar{b}= 0.430\nonumber\\&
\tilde{A} = 0.070, \tilde{b} = -0.250, \bar{x} = 0.102\nonumber\\&
A_g = 168, C_g = 2.475, c_g = 0.765, r = 0.00116, u = 0.001
\label{eq.14}
\end{align}

\begin{figure}[h]
\centering
\resizebox{0.48\hsize}{!}
{\includegraphics*{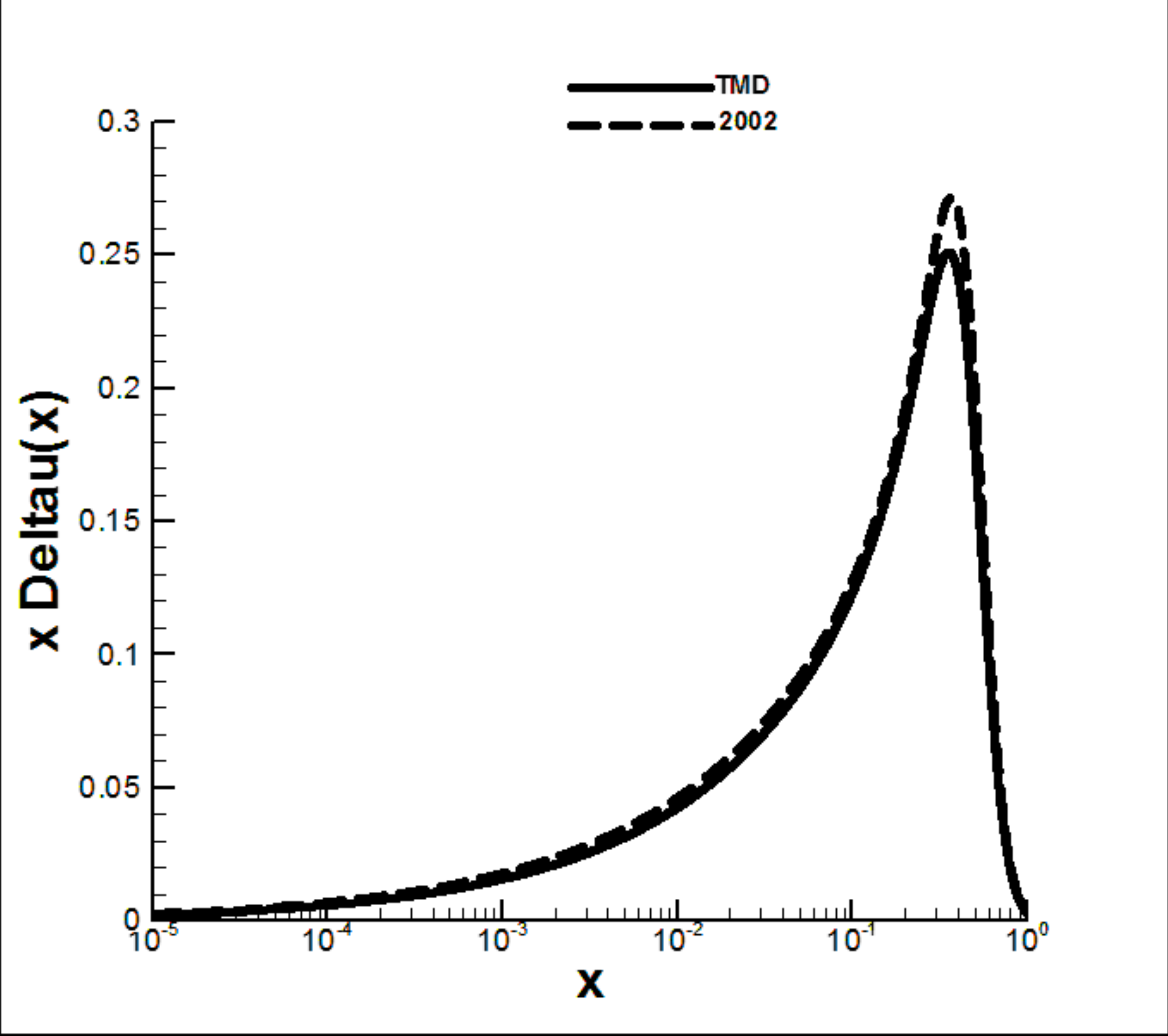}}
\resizebox{0.48\hsize}{!}
{\includegraphics*{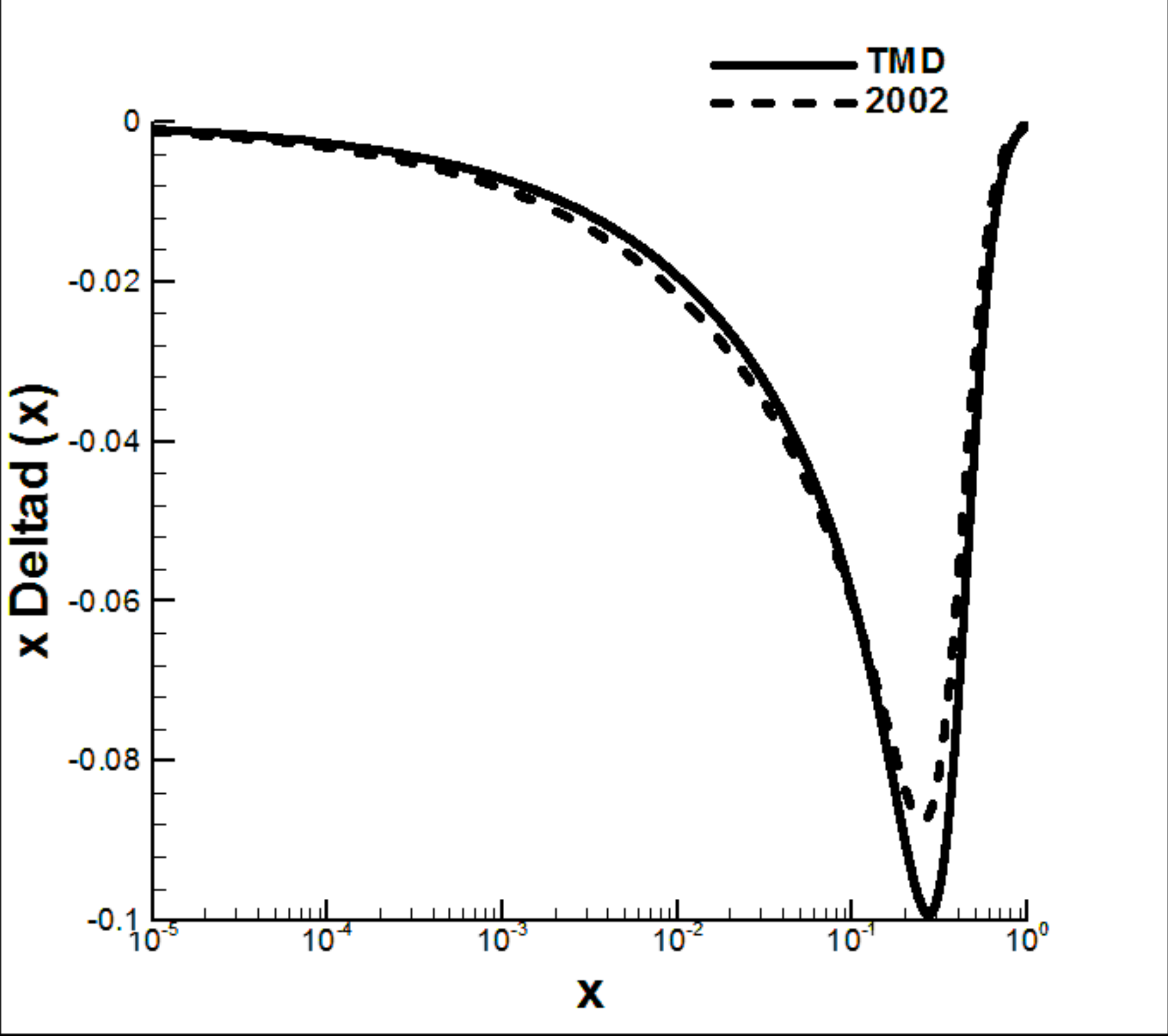}}
\caption{The distribution of $\Delta {u}$ and $\Delta {d}$ in comparison to the 2002 results.}
\label{fig3}
\end{figure}

The excellent agreement shown in these figures
and the agreement of the parameters with the ones found in [6] with the only exception of $\tilde{A}$ fixed by data at small $x$ changed since 2002 (the
exponent for the non-diffractive term for the light anti-quarks has been
chosen to be equal to the one for the valence partons) is a good
confirm of our approach, which also implies the exponential behavior
$\emph{exp}{(-x/\bar{x})}$ at high $x$ shown experimentally by the distributions.
As long for the gluons, the Planck expression reproduces the small $x$ behavior
and the exponential fall at large $x$ of $x G(x)$, but
needs the "ad hoc" factors introduced in (11) to get the agreement with data.
The parameters found are in a very good agreement with the ones
found in [6], as shown by the comparison of the parameters:

\begin{center}
\begin{tabular}{l|*{3}{|c}r}
               & ref. [6]  & this paper  & $\frac{A^{\prime}}{A} ln{(1 + \exp{\tilde{Y}_q})}$& \\
\hline
$\tilde{X}_{u_+}$               & 0.46188 & 0.446   & 0.4650\\
$\tilde{X}_{d_-}$               & 0.30174 & 0.320   & 0.3115\\
$\tilde{X}_{u_-}$               & 0.29766 & 0.297   & 0.2975\\
$\tilde{X}_{d_+}$               & 0.22775 & 0.222   & 0.2345\\
$b$                             & 0.40962 & 0.43   & \\
$\tilde{A}$                     & 0.08318 & 0.070   & \\
$\tilde{b}$                     & -0.25347 & -0.250   & \\
$\bar{x}$                       & 0.09907 & 0.102  & \\
\end{tabular}
\end{center}

where we have reported in the third column the
coefficients found with the extension to the transverse
momenta, comparing them with the "ad hoc" factors
$\tilde{X}_q$ introduced in [6].

\begin{figure}[h]
\centering
\resizebox{0.48\hsize}{!}
{\includegraphics*{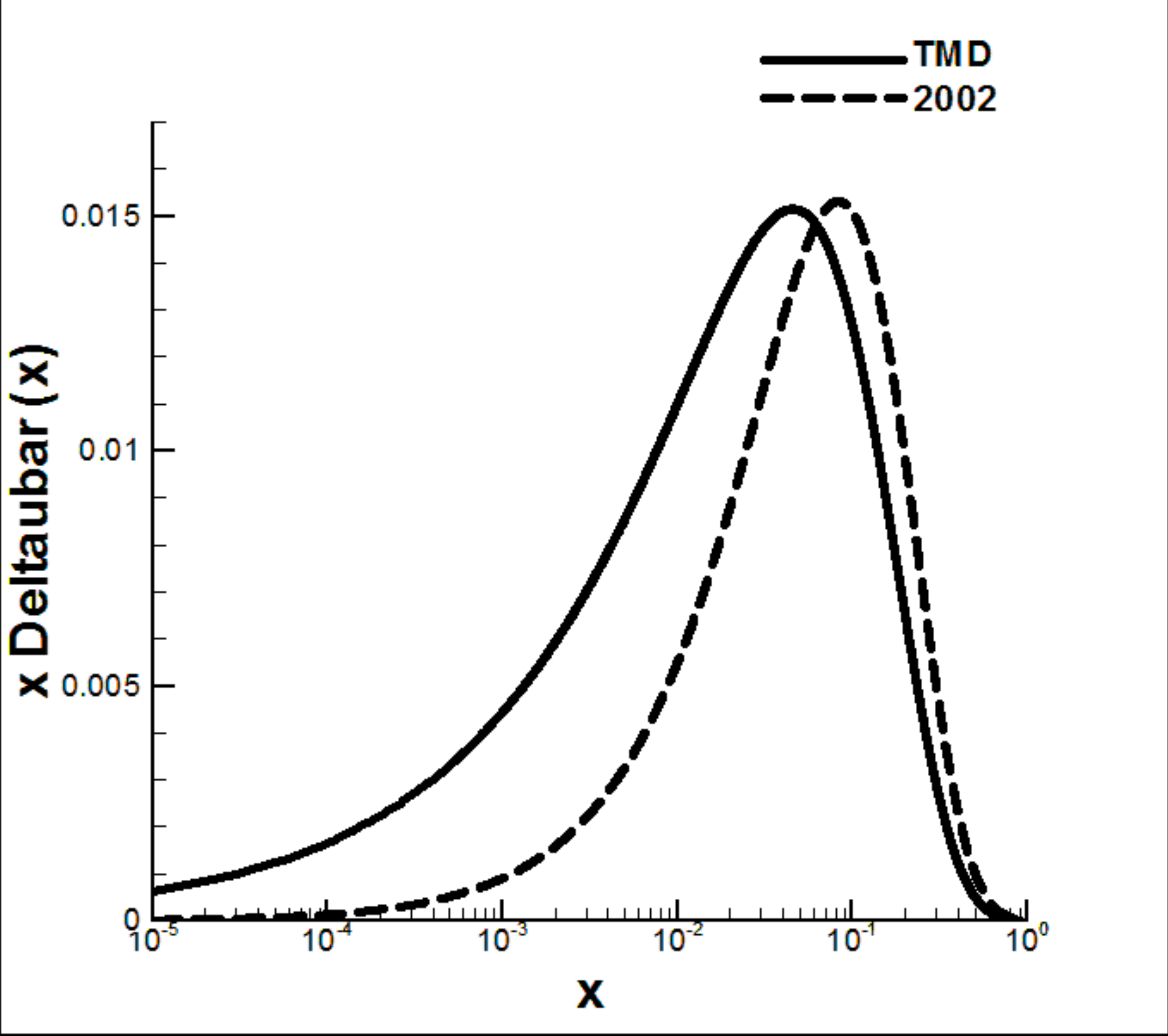}}
\resizebox{0.48\hsize}{!}
{\includegraphics*{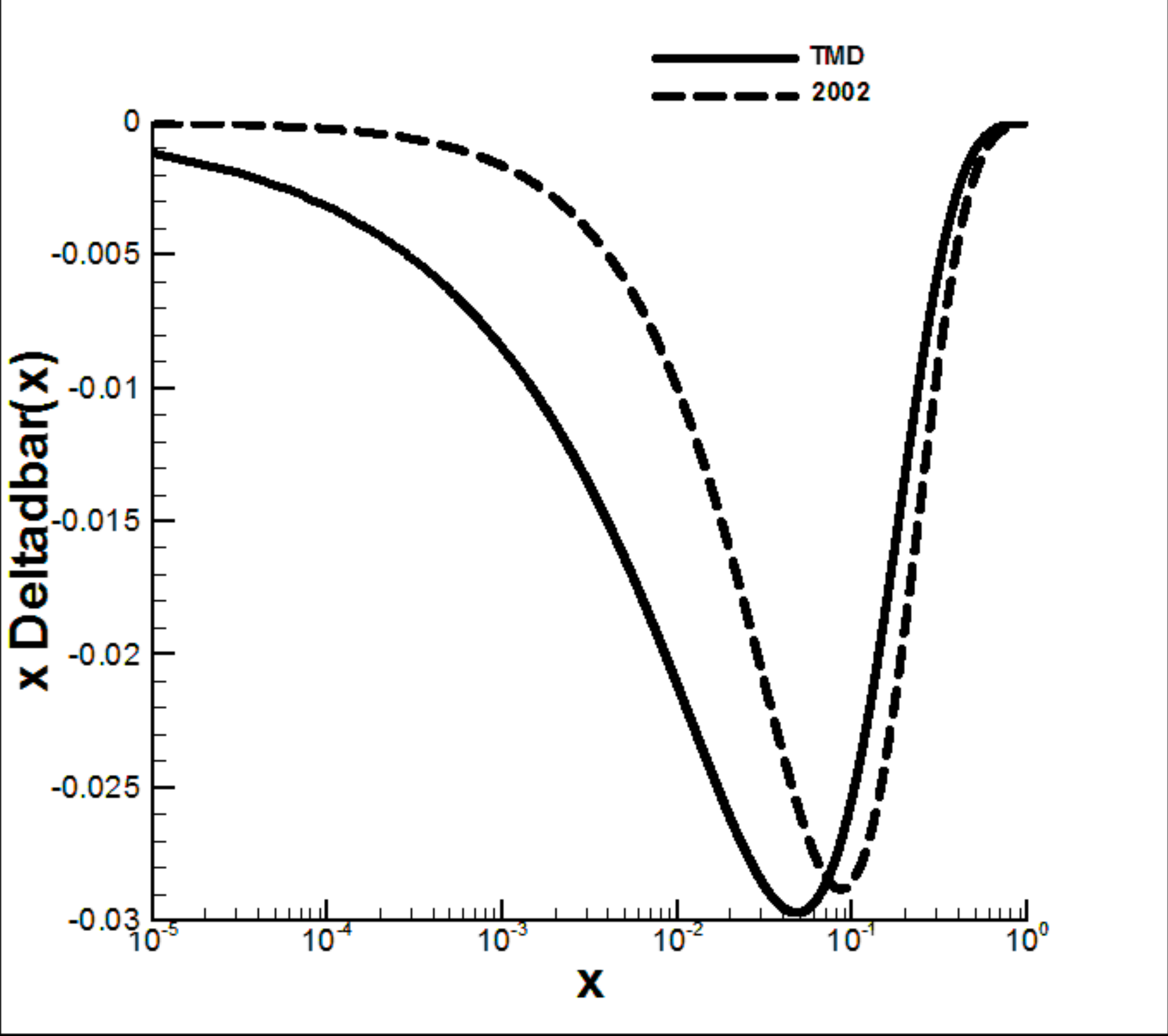}}
\caption{The distribution of $\Delta {\bar{u}}$ and $\Delta {\bar{d}}$ in comparison to the 2002 results.}
\label{fig4}
\end{figure}

\begin{figure}[h]
\centering
\resizebox{0.6\hsize}{!}
{\includegraphics*{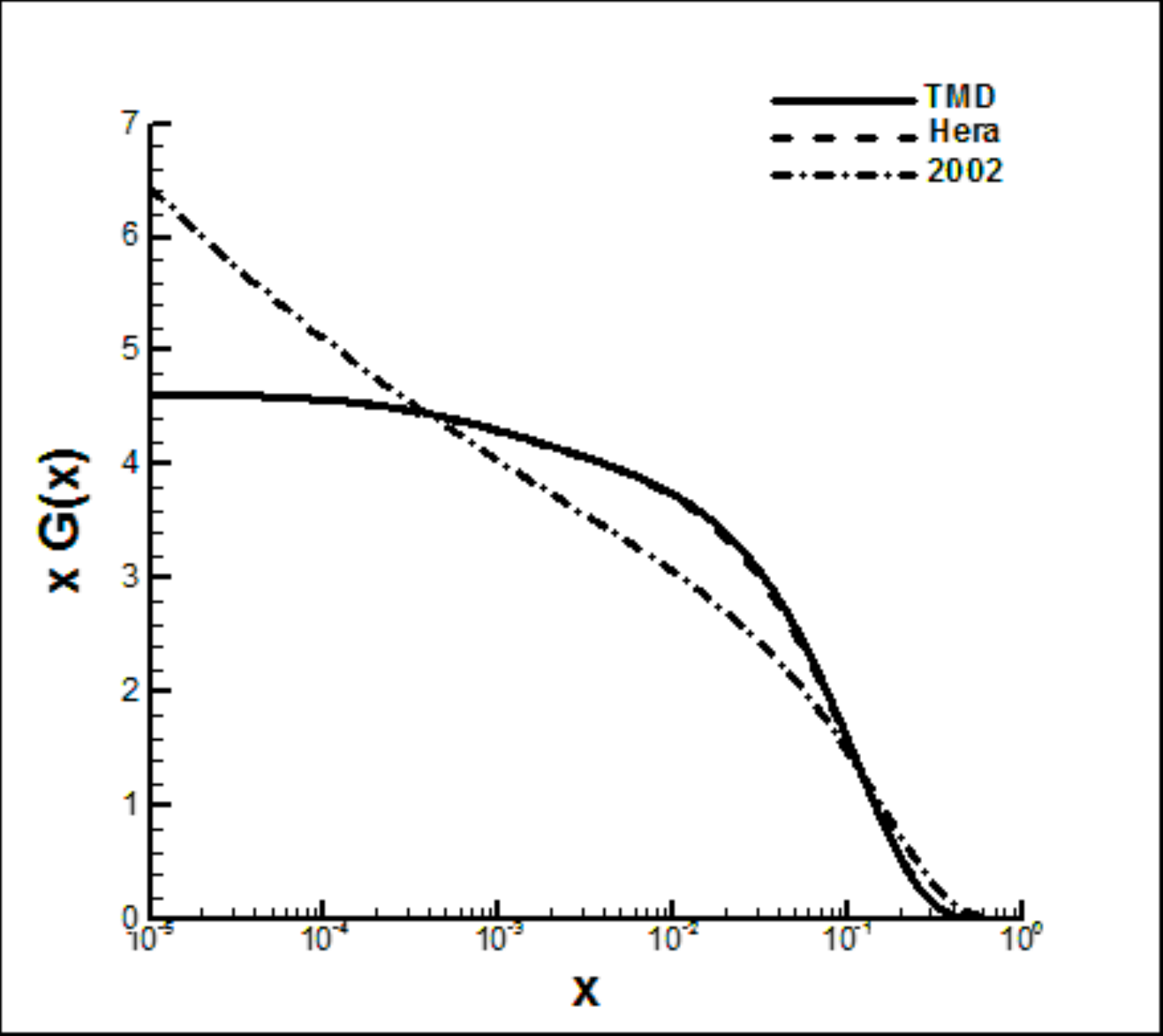}}
\caption{The distribution of gluons in comparison to the 2002 results and Hera data.}
\label{fig5}
\end{figure}

In figures 6-7-8 the ratios
$\frac{d(x)}{u(x)}$, $\frac{\Delta u(x)}{u(x)}$ and
$\frac{\Delta d(x)}{d(x)}$ are compared with [6] and for the first
of them also with Hera. Finally in fig.9
we compare $x\bar{d}(x) - x\bar{u}(x)$ with the expression found in
[6]. The exponential behaviour $exp{(-x/\bar{x})}$
predicted by the statistical model (for the $\bar{q}$'s with negative
$\tilde{X}$'s with modulus larger than $\bar{x}$ the Boltzmann limit is a good
approximation) agrees with the data from the Fermilab
E866 Drell-Yan experiment [14] displayed in [15], where the
definitions "intrinsic" and "extrinsic" correspond here to "non diffractive"
and "diffractive" terms.

\begin{figure}[h]
\centering
\resizebox{0.6\hsize}{!}
{\includegraphics*{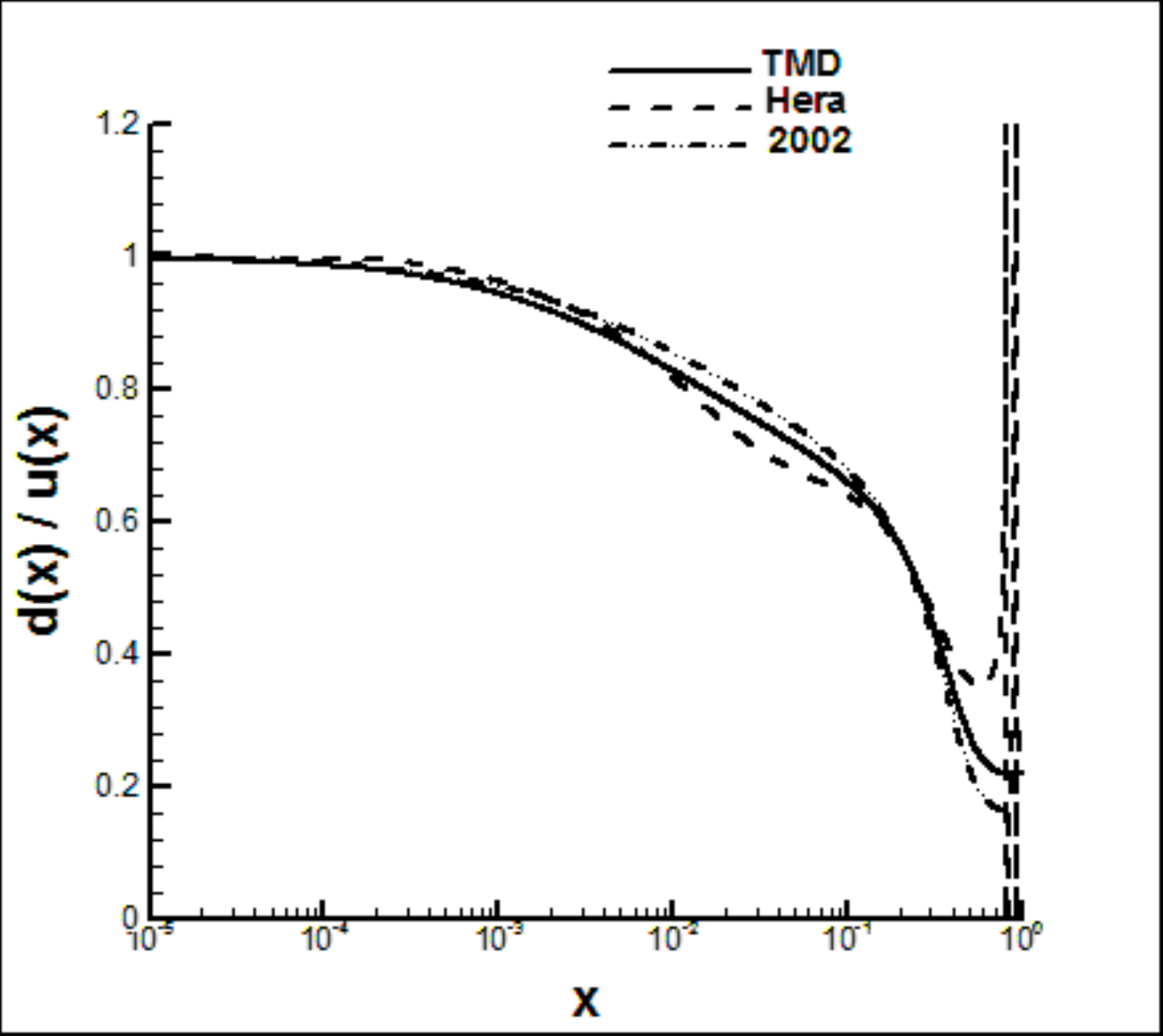}}
\caption{The distribution of $d(x)/u(x)$ in comparison to the 2002 results and Hera data.}
\label{fig6}
\end{figure}

\begin{figure}[h]
\centering
\resizebox{0.6\hsize}{!}
{\includegraphics*{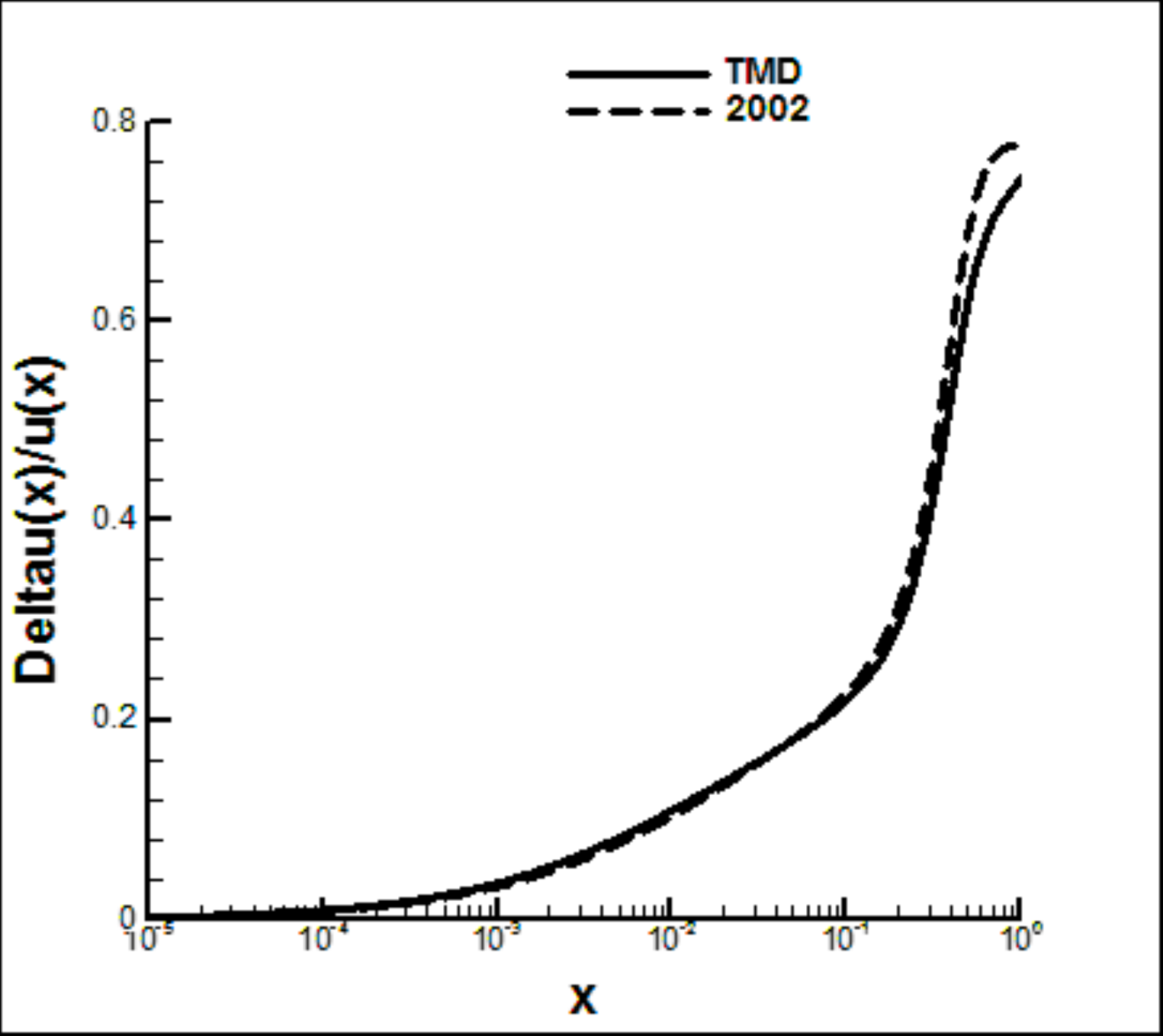}}
\caption{The distribution of $\Delta{u(x)}/u(x)$ in comparison to the 2002 result.}
\label{fig7}
\end{figure}

\begin{figure}[h]
\centering
\resizebox{0.6\hsize}{!}
{\includegraphics*{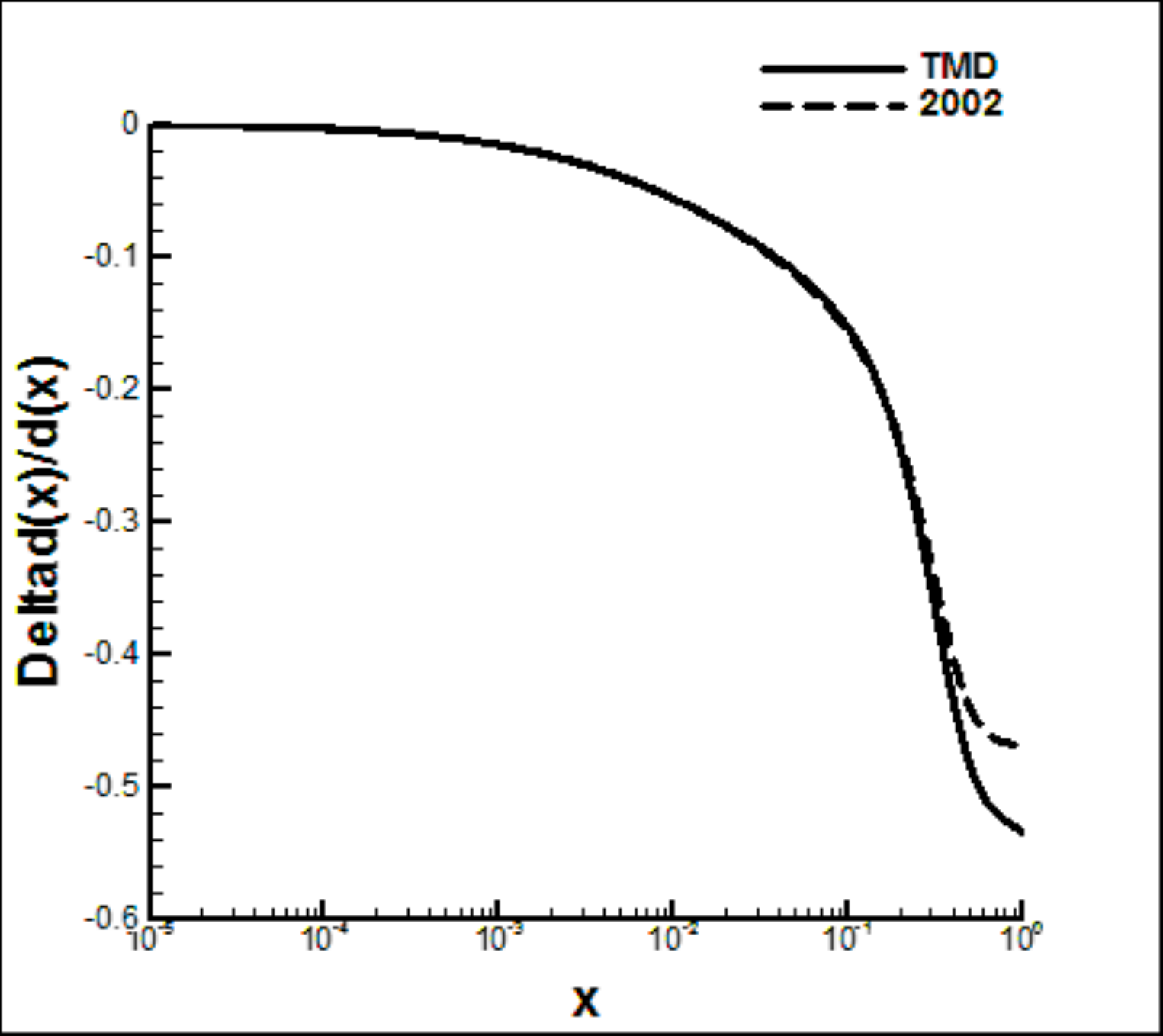}}
\caption{The distribution of $\Delta{d(x)}/d(x)$ in comparison to the 2002
result.}
\label{fig8}
\end{figure}

\begin{figure}[h]
\centering
\resizebox{0.6\hsize}{!}
{\includegraphics*{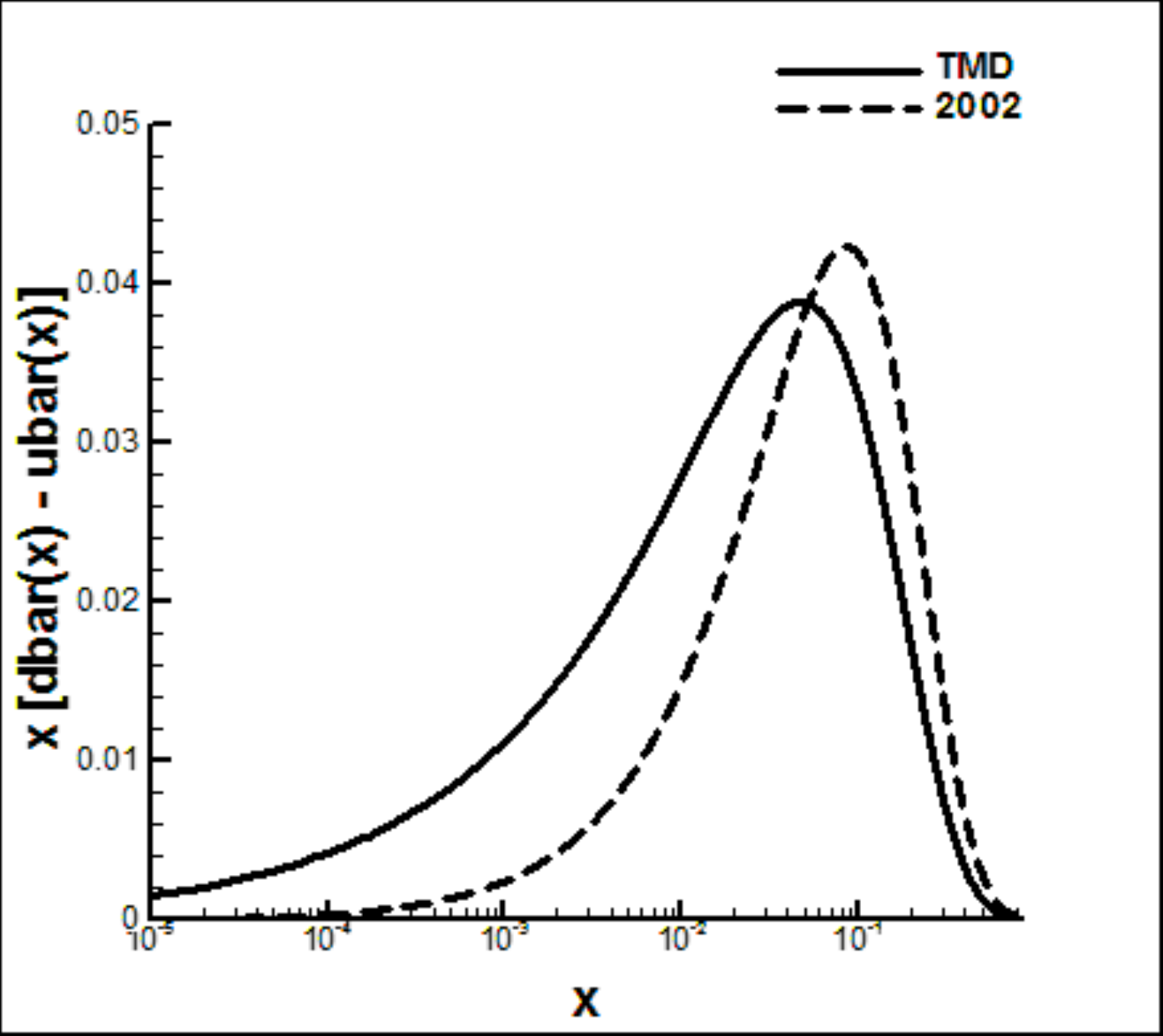}}
\caption{The distribution of $x\bar{d}(x)- x\bar{u}(x)$ in comparison to the 2002
result.}
\label{fig9}
\end{figure}

\begin{center}
\begin{tabular}{l|*{3}{|c}r}
               & Hera LSR &   this paper LSR &   this paper TSR  \\

\hline
$Gluon$           & 0.410 & 0.409 & 0.4265  \\
$u$               & 0.312 & 0.298 & 0.3153   \\
$d$               & 0.162 & 0.152 & 0.1758   \\
$\bar{u}$         & 0.0313 & 0.0243 & 0.0145   \\
$\bar{d}$         & 0.0366 & 0.0322 & 0.0377   \\
\end{tabular}
\end{center}

\begin{center}
\begin{tabular}{l|*{2}{|c}r}
               & ref. [6]  & this paper  & \\
\hline
$\int{[\Delta{u}(x)]dx}$        & 0.663811 & 0.642376  \\
$\int{[\Delta{d}(x)]dx}$           & -0.255714 & -0.258968   \\
$\int{[\Delta{\bar{u}}(x)]dx}$       & 0.0464154 & 0.0671633   \\
$\int{[\Delta{\bar{d}}(x)]dx}$      & -0.0865359 & -0.13182   \\
$\int{[\bar{d}(x)-\bar{u}(x)]dx}$    & 0.12739  & 0.17123   \\
\end{tabular}
\end{center}

\begin{center}
\begin{tabular}{l|*{4}{|c}r}
               & Hera &   this paper &   ref. [6] &   ref. [13]  \\

\hline
$d(1)/u(1)$       & 0.31167 & 0.219354 & 0.159891 & 0.22 \\
$\Delta{u(1)}/u(1)$  &     & 0.742825  & 0.77773  & \\
$\Delta{d(1)}/d(1)$  &     & -0.53318  & -0.467774  & \\
\end{tabular}
\end{center}

In the Tables we compare the contribution to the longitudinal and transverse sum rules of the fermion and of the gluon with Hera result
for the longitudinal sum rule and the first moments of the polarized parton distributions with [6]. The ratios of the
limits of the parton distributions at $x=1$ are also shown
and compared with [6], and also with [10] and [13] for
$d(1)/u(1)$.
A point in favour with our parametrization with the universal Boltzmann
behaviour $\emph{exp}{(-x/\bar{x})}$ at high $x$ is the fact that, despite we
fixed the parameters for the unpolarized distributions in order to be as much
as possible equal to [10], in the $x \rightarrow 1$ limit, where their
extrapolation is less predictive, the ratio $d(1)/u(1)$ is in better
agreement with [13] for the statistical distributions.
With respect to [16], while
$\frac{\Delta u(1)}{u(1)} + \frac{4 d(1)}{3 u(1)}$ is near to
$1$, the value expected in that paper, for $\Delta d(1)/d(1)$
we predict a value more negative than $\frac{-1}{3}$.
The sum of the numbers in the third column of the second table should give $ 1 $ to obey the transverse energy sum rule (which is divided by $M^2$).

\label{sec:3}

\section{Conclusion}

The purpose of this paper was to perform a check-up of the statistical
parton model by taking advantage of our understanding of the transverse
distributions, which improves the theoretical consistency of our statistical
approach, through the comparison with the Hera fit based on the combined
analysis of $H_1-ZEUS$ data performed after ref.[6]. The fact that we
have required the polarized distributions equal [6] is
fully justified by the remarkable success in describing polarized data
again taken after both for $g_1^{ p, d, He_3}$
[7, 8] and for the more recent $W^{\pm}$ production[12].
The reader may judge by looking to the figures, we present, with the
values of the parameters found so similar to the ones proposed in 2002
on the goodness of that proposal, which in our judgement may help in
getting from experiment the parton distributions, in particular the
polarized ones, which we have been able to well describe in their
qualitative and quantitative properties. The choice of the same exponent, b, for the exponent of the power factor
in the non diffractive part of the fermion parton distributions is well
consistent with the Hera result[10] for the unpolarized light anti-quark
distributions and implies larger polarizations  for them at low $x$ with
respect to [6] to be compared with experiment in that region, a more positive contribution to the Bjorken sum rule [17] and a larger negative contribution to the Gottfried
sum rule [2].
As long as for gluons the Planck form $x G(x)$ proportional to $\frac{x}{e^{x/\bar{x}} - 1}$
looks like the Hera result with a rather flat behavior at small $x$ and
the exponential behavior at high $x$ with the same exponent of the valence partons, but to coincide with it needs the slowly varying factor written
in eq.(11), for which at the moment we are not able to give an interpretation.
The $p_T$ dependance implied by the transverse sum rule
for $p^2_T$ larger than $\mu^2 x \tilde{Y}_q$ approaches a
gaussian with width $ \mu \sqrt{x}$, proportional to $\sqrt{x}$,
as in [18]. In the classical limit,
neglecting $\frac{p^2_T}{P^2_z}$ and the power
dependance on $x$ and $p_T$, by integrating first
in $x$ with the gaussian approximation for the exponential,
one gets the behavior $\emph{exp}{(\frac{- 2 p_T}{\mu \sqrt{\bar{x}}})}$
[19], with an "effective temperature"
$\frac{\mu \sqrt{\bar{x}}}{2} = 49 (MeV)$ smaller than the range
$120-150 (MeV)$ proposed in [20], but the effect of of quantum statistics leads
to a harder spectrum for $p_T$, as it happens for $x$, since the local
maximum for $x q(x)$ for the valence partons is larger than $\bar{x}$,
as it is shown in figures 1 and 2. In fact for the non diffractive contribution of the valence partons,
which mainly contribute to the larger $p_T$, we have:
\begin{equation}
(\overline{{p}_T})_u = 97 (MeV)\nonumber
\end{equation}
\begin{equation}
(\overline{{p}_T})_d = 81 (MeV)\nonumber
\end{equation}

both larger than $49 (MeV)$.\\

It is worth to stress the attractive feature of the
quantum statistical distributions of fixing the free
parameters in regions of $x$, where data have a larger
statistics and small systematic errors.
In fact, while $A^{\prime}$, $\bar{A}^{\prime}$ and $A_G$ are constrained
by eqs.(7a-7b) and by the condition that parton
carry the proton momentum, $\tilde{A}$ and $\tilde{b}$ are fixed
by the measurements at small $x$, $\bar{x}$, the
$\tilde{X}$'s and the $\tilde{Y}$'s are fixed by the
comparison with the intermediate region, where the valence
quarks dominate, and determine the normalization and the
Boltzmann behaviour $\emph{exp}{(-x/\bar{x})}$, where the data
are scarce and in particular $F^n_2(x)$ is difficult to extract
from $F^d_2(x)$ as a consequence of the Fermi motion.
Also the disentangling of the $q$ and $\bar{q}$ contributions
to the e. m. DIS, which for the unpolarized distributions is
achieved with the help of eqs. (7a-7b), implied by the equilibrium
conditions (4) and (5) brings to the prediction of a positive (negative)
$\Delta \bar{u}(x)$ ($\Delta \bar{d}(x)$) in agreement [11] with the
measurement of $W^{\pm}$ at RHIC.
Also the exponential behaviour of the isospin and spin asymmetries
of the sea with the same slope of the high x behaviour of the valence
partons and of the gluons provides a good test for the statistical
distributions.\\

\textbf{AKNOWLEDGEMENTS}

We are grateful to Professor Jacques Soffer, who first realized
the strong similarity between the statistical distributions and
the ones found at Hera, which gave the inspiration for this work. One of us (F. B.) is very grateful to the organizers of HIX2014,
since his participation to this conference allowed him to
have fruitful discussions with Professors Alberto Accardi, Chieh Peng,
Anthony Thomas and Werner Vogelsang.

\end{document}